\documentclass[10pt, conference, compsocconf]{IEEEtran}
\usepackage{graphicx}
\usepackage{booktabs}
\usepackage{subcaption}
\usepackage{caption}
\usepackage{multirow}
\usepackage{amsmath}
\usepackage{lscape}
\usepackage{mathtools}
\usepackage{amsfonts}
\usepackage{graphicx}
\DeclareGraphicsExtensions{.png}

\usepackage{multirow}
\input{Qcircuit.tex}

\ifCLASSINFOpdf
\else
\fi
\begin{document}
\title{Design of Quantum Circuits for Galois Field Squaring and Exponentiation}


\author{\IEEEauthorblockN{Edgard Mu\~{n}oz-Coreas and Himanshu Thapliyal}
\IEEEauthorblockA{Department of Electrical and Computer Engineering \\
University of Kentucky \\
 Lexington, KY \\
Email: hthapliyal@uky.edu}
}

    \IEEEoverridecommandlockouts
    \IEEEpubid{\makebox[\columnwidth]{978-1-5090-6762-6/17/\$31.00  \copyright 2017 IEEE \hfill} \hspace{\columnsep}\makebox[\columnwidth]{ }}

\maketitle

\begin{abstract}

This work presents an algorithm to generate depth, quantum gate and qubit 
optimized circuits for $GF(2^m)$ squaring in the polynomial basis.  
Further, to the best of our knowledge the proposed quantum squaring circuit algorithm is the only work that
considers depth as a metric to be optimized.  We 
compared circuits generated by our proposed algorithm against the state of the 
art and determine that they require $50 \%$ fewer qubits and offer gates 
savings that range from $37 \%$ to $68 \%$. Further, existing quantum exponentiation are based on either modular or integer arithmetic.  However, Galois arithmetic is a useful tool to design resource efficient quantum exponentiation circuit applicable in quantum cryptanalysis. 
Therefore, we present the quantum circuit implementation of Galois field exponentiation based on the proposed quantum Galois field squaring circuit. 
We calculated a qubit savings ranging between  $44\%$  to $50\%$  and quantum gate savings ranging between $37 \%$ to 
$68 \%$ compared to identical quantum exponentiation circuit based on existing squaring 
circuits. 
\end{abstract}


% \fi
\IEEEpeerreviewmaketitle

\section{Introduction}

\label{motivation_and_purposeeeeeeeeeee}
%\subsection{Paper Organization} 

Among the emerging computing paradigms, quantum computing appears to be promising due to its applications in number 
theory, cryptography, search and scientific computation \cite{Cheung} \cite{Polian_DAC}.  Quantum computing is also being investigated for its promising application in high performance computing \cite{Cheung} \cite{Polian_DAC}.  Quantum circuits do not lose information during computation and quantum computation can only be performed when the system consists of quantum gates.  Quantum circuits generate a unique output vector for each input vector, that is, there is a one-to-one mapping between the input and output vectors.  Any constant inputs in the quantum circuit are called ancillae and garbage output refers to any output which exists in the quantum circuit to preserve one-to-one mapping but is not one of the primary inputs nor a useful output.  The inputs regenerated at the outputs are not considered garbage outputs \cite{Fredkin}.

 Fault tolerant implementation of quantum circuits is gaining the attention of researchers because physical quantum computers are prone to noise errors \cite{Polian_DAC}.  Fault tolerant implementations of quantum gates and quantum error correcting codes can be used to overcome the limits imposed by noise errors in implementing quantum computing \cite{Polian_DAC}.  Recently, researchers have implemented quantum logic gates and circuits such as the controlled phase gate, controlled square-root-of-not gate, Toffoli gate, Fredkin gate and quantum full adders with the fault tolerant \textit{Clifford+T} gate set due to its demonstrated tolerance to noise errors \cite{Miller}.
 
 A Galois field is a set that possesses a particular set of mathematical properties. Galois fields and Galois field arithmetic have drawn interest of many 
researchers because of its applications in encryption, error correcting codes 
and signal processing \cite{Paar} \cite{Pruss-DAC}.  Galois field squaring has drawn interest because it is a lower cost alternative to multiplication when both operands are the same.
  There are existing designs of quantum circuits for Galois field squaring such as the designs in \cite{S1}.  The Galois field squaring designs in \cite{S1} are 2 input and 2 output designs.  They have the mapping $(A,B)$ to $(A,A^2 + B)$.  To implement $A^2$, $B$ is replaced by ancillae.  In a quantum circuit, the ancillae and garbage outputs are considered overhead and need to be minimized. This is  because an increase in number of ancillae inputs and garbage outputs will increase the number of qubits in the quantum circuit.  Therefore, in applications that require repeated squaring operations such as Galois field exponentiation, the use of the  existing quantum squaring circuit design in \cite{S1} will result in significant ancillae overhead.  

\textit{In this work we present an algorithm to realize depth, qubit and quantum gate optimized circuits for Galois field squaring.  Further, to the best of our knowledge the proposed quantum squaring circuit algorithm is the only work that considers depth as a metric to be optimized. In this work, depth is defined as the number of parallel sets of 
quantum gates in a circuit.  Quantum gates that operate on independent sets of 
qubits are considered parallel.  The quantum Galois field squaring circuits are based on the Feynman gate.  The Feynman gate is a member of the fault tolerant \textit{Clifford+T} gate set \cite{Miller}.}  The quantum squaring circuit based on our proposed algorithm is compared and is shown to be better than the existing design of quantum Galois field squaring circuits in \cite{S1}. The comparison is performed in terms of number of gates and qubits. 

Further, existing quantum exponentiation are based on either modular or integer arithmetic. However, quantum exponentiation based on Galois arithmetic is applicable in quantum cryptanalysis \cite{Cheung}.  \textit{Therefore, we present the quantum circuit implementation of Galois field exponentiation based on the proposed quantum Galois field squaring circuit.}  We calculated a significant  qubit and quantum gate savings compared to identical quantum exponentiation circuit based on existing squaring circuit in \cite{S1}.

%Furthermore, we present an algorithm to realize Galois field exponentiation as a quantum circuit.  The quantum Galois field exponentiation circuit is based on a quantum Galois field multiplier and Galois field squarer.  Quantum exponentiation circuits made with quantum circuits from our proposed quantum Galois field squaring circuit are compared and shown to be better than quantum exponentiation circuits made with existing designs of quantum Galois field squaring circuits in terms of number of gates and qubits. 
    
The paper is organized as follows: the basics of Galois fields is presented in section \ref{backgroundddddd}; the proposed Galois field squaring circuit algorithm and analysis are presented in sections \ref{algorithmmmmm} and \ref{squarzanalytics}. The proposed Galois field exponentiation circuit algorithm and analysis are  presented in sections \ref{expo} and \ref{expoanalytics}, while sections \ref{squarzperformance}, \ref{expoperformance} and \ref{finis} provide the comparison of the Galois field squaring circuits, Galois field exponentiation circuits and conclusions, respectively.

\section{Basics of Galois Fields}
\label{backgroundddddd}

%Galois fields and Galois field arithmetic have drawn interest of many researchers because of its applications in encryption, error correcting codes and signal processing \cite{Paar} \cite{Pruss-DAC}.  

A Galois 
field is a set that possesses a particular set of mathematical properties.  The number of elements in the field is determined by the 
field basis ($f(x)$).  \textit{In our work, we represent $f(x)$ with 
irreducible polynomials.}  The set of elements in the Galois field will 
contain all possible polynomials that are less than the basis $f(x)$.  A field 
element is given as:

%Show equation here
\begin{equation}
\alpha_{m-1} \cdot x^{m-1} + \alpha_{m-2} \cdot x^{m-2} + \cdots + \alpha_{1} 
\cdot x + \alpha_{0} 
\label{equation:GFS_DAC_1}
\end{equation}

Where $\alpha \in GF(2)$ and $\{0,1\} \in GF(2)$.

Galois field applications make use of the Galois field addition, multiplication and exponentiation operations. Next, we discuss these important Galois field arithmetic operations.

\subsection{Galois field arithmetic}

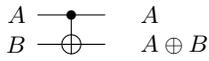
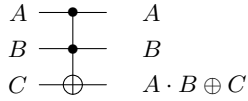
\begin{figure}
	\small
	\flushleft
	\begin{subfigure}[hb]{1.5in}
	\flushleft
		\[
		\Qcircuit @C=1em @R=.7em {
		\lstick{A} & \ctrl{1} & \qw & \rstick{A}\\
		\lstick{B} & \targ & \qw & \rstick{A \oplus B} \\
		} 
		\]
	\caption{Feynman Gate}
	\end{subfigure}  \qquad \begin{subfigure}[hb]{1.5in}
	\flushleft
		\[
		\Qcircuit @C = 1em @R = .7em @!R{
		\lstick{A} & \ctrl{1} & \qw & \rstick{A}\\
		\lstick{B} & \ctrl{1} & \qw & \rstick{B}\\
		\lstick{C} & \targ & \qw & \rstick{A \cdot B \oplus C} \\
		}
		\]
	\caption{Toffoli Gate}	
	\end{subfigure}
	\caption{Gates used in quantum circuits for Galois field arithmetic.}
\label{figure:GFS1}
\end{figure}

Addition in Galois fields is the bit-wise exclusive-OR (XOR) operation.  In quantum computing, the bit-wise XOR operation is implemented with the Feynman gate.  Figure \ref{figure:GFS1} shows the schematic symbol of the Feynman gate.

Multiplication in Galois fields evaluates the following equation:

\begin{equation}
	Y = A \cdot B \mod f(x)
	\label{equation:GFS_DAC_2}	      
\end{equation}

Where $f(x)$ is the field basis and ${A,B} \in GF(2^m) $.  In the most straight-forward implementation, partial products are produced with the AND operation.  All addition is Galois and it along with the reduction modulo $f(x)$ is performed with the XOR operation.  In quantum computing, the AND operation is implemented with the Toffoli gate.  Figure \ref{figure:GFS1} shows the schematic symbol of the Toffoli gate. 

Squaring is a special case of multiplication where both operands are the same. Squaring in Galois fields evaluates the following equation \cite{Paar}:

\begin{equation}
A^2 = \sum_{i = 0}^{n-1} \alpha_i \cdot x^{2 \cdot i} \mod f(x)
\label{equation:GFS_DAC_3}
\end{equation}  

Where $n$ is the order of the field basis $f(x)$ and $A \in GF(2^m)$. Equation \ref{equation:GFS_DAC_3} is implemented in hardware with the XOR operation.

Exponentiation is performed with repeated applications of the Galois field squaring and Galois field multiplication operations.  \textit{In our work, we are interested in quantum circuits that realize the following exponential equation:}

\begin{equation}
	Y = A^{2 ^ n-2} \mod f(x) 
	 \label{equation:GFS_DAC_4}
\end{equation}	

Where $n$ is the order of the field basis $f(x)$ and $A \in GF(2^m)$.  
Equation \ref{equation:GFS_DAC_4} has drawn interest of researchers because 
equation \ref{equation:GFS_DAC_4} is an implementation of the Galois field 
inversion operation \cite{Wu}.  The literature shows that 
polynomial basis implementations of equation \ref{equation:GFS_DAC_4} can 
excel against alternative Galois field basis representations in terms of total 
bit operations \cite{Wu}.

\section{Proposed Galois Field Quantum Squaring Algorithm} 
\label{algorithmmmmm}

The quantum Galois field squaring circuit is designed with less qubit and quantum gate cost compared to existing 
Galois field squaring circuit design approaches \cite{S1}.  Further, the proposed algorithm includes circuit depth as an 
optimization criterion.  Consider the squaring of one $n$ bit field number $A$ 
stored at quantum register $\ket{A}$.  At the end of the computation, the 
quantum register $\ket{A}$ will have $\ket{Y}$ (where $Y = A^2 \mod f(x)$ and 
$f(x)$ is the field basis).

\begin{figure*}[t!bhp]
\centering
	\includegraphics[width = 5.5in]{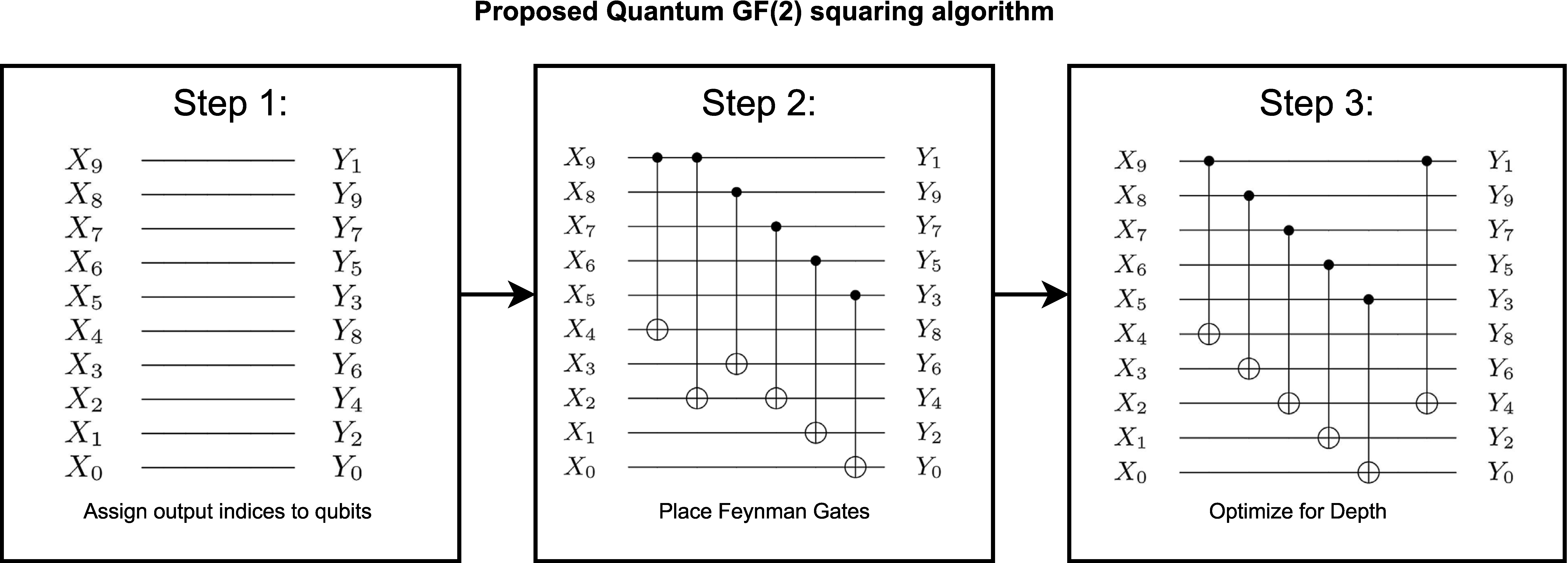}
	\caption{Generation of quantum Galois field squaring circuit for the field with basis $f(x) = x^{10}+x^3 + 1$: Steps 1-3.}
\label{GSF_DAC_fig:1}	
\end{figure*}

The generalized algorithm of designing a Galois field squaring circuit is 
explained below along with an illustrative example in figure 
\ref{GSF_DAC_fig:1} of a Galois field squaring circuit for a field of size 
$10$ defined with basis $f(x) = x^{10}+x^3 + 1$ that can perform the squaring 
of any field element.  Details of the proposed algorithm are discussed below:

\subsection{Steps of Proposed Algorithm for Galois Field Squaring Circuits}

\begin{itemize}

\item Step 1: For $i = 0:1:n-1$: Evaluate the following:
\begin{equation}
	\alpha_i \cdot x^{2 \cdot i} \mod f(x)
	\label{equation:GFS_DAC_8}
\end{equation}

If equation \ref{equation:GFS_DAC_8} equals $\alpha_i \cdot x^{2 \cdot i}$, then at the 
end of 
computation the qubit $\ket{\alpha_i}$ will have the value $\ket{Y_{2 \cdot i}}$.  
Else, equation \ref{equation:GFS_DAC_8} equals the following:

\begin{equation}
	B_{n-1} \cdot x^{n-1} + B_{n-2} \cdot x^{n-2} + \cdots + B_{1} \cdot x + 
	B_0  
	\label{equation:GFS_DAC_9}
\end{equation}

Where $B_j \in {0,1}$ and $0 \leq j \leq n-1$.

At the end of computation the qubit $\ket{\alpha_i}$ will have the value 
$\ket{Y_{j}}$ provided  $B_j = 1$ and $\ket{Y_{j}}$ is not already assigned to 
another qubit in $\ket{A}$.  

\end{itemize}
Thus, the values in $\ket{Y}$ will not be in sequential order but the generated squaring circuits will require no ancillae.  
The existing quantum squaring circuit design in \cite{S1} requires $n$ ancillae.  Thus, the proposed circuits reduce qubit cost by $50 \% $.

%The quantum Galois field squaring circuit for the Galois field with basis $f(x) = x^{10}+x^3 + 1$ after this Step is shown in figure \ref{GSF_DAC_fig:1}. 

\begin{itemize}

\item Step 2: Repeat Step 2 For $i = \lceil \frac{n-1}{2} \rceil:1:n-1$:

For $k = 0:1:n-1$:

If $B_k = 1$ in equation \ref{equation:GFS_DAC_9} for qubit $\ket{A_i}$ and qubit $\ket{A_i}$ will not have the value $\ket{Y_{k}}$, apply a Feynman gate such that $\ket{A_i}$ and the qubit that will have the value $\ket{Y_{k}}$ are passed to the inputs $A$ and $B$ respectively. 	
	
  %The quantum Galois field squaring circuit for the Galois field with basis $f(x) = x^{10}+x^3 + 1$ after this Step is shown in figure \ref{GSF_DAC_fig:1}.

\item Step 3: This Step has three sub-steps:
\begin{itemize}

\item Step 1: select a gate as a reference gate
\item Step 2: If a gate shares a qubit with reference gate, move it to the end of the circuit
\item Step 3: If more than 1 gate is moved then repeat Steps 1 and 2 with only the moved gates 
\end{itemize}
%Reorganize the Feynman gates to minimize the circuit depth.  By maximizing the number of Feynman gates acting in parallel, the depth can be optimized. 

%We use a sorting algorithm to place Feynman gates in parallel.  

%The sorting algorithm passes through the circuit and .  This process is repeated until       
%The quantum Galois field squaring circuit for the Galois field with basis $f(x) = x^{10}+x^3 + 1$ after this Step is shown in figure \ref{GSF_DAC_fig:1}.  The depth of the circuit in figure \ref{GSF_DAC_fig:1} has been reduced by one.

\end{itemize}

\textbf{Theorem:} \textit{ Let $A$ be a $n$ bit binary number and $A \in 
GF(2^n)$, then the steps in the proposed Galois field squaring circuit 
algorithm results in a quantum Galois field squaring circuit that functions 
correctly.  The proposed algorithm designs an $n$ bit Galois field squaring 
circuit that produces the output $A^2$ on the quantum register where $A$ is 
originally stored.}

\textbf{Proof:} The proposed algorithm will make the following changes on the 
input:

\begin{itemize}

\item Step 1:  Step 1 of the proposed algorithm transforms the input states to

\begin{equation}
	\ket{ \sum_{i=0}^{\lfloor \frac{n-1}{2} \rfloor}   A_i \cdot x^{2 \cdot i} 
	\mod f(x)}
	\label{equation:GFS_DAC_6}
\end{equation}
    
Where $f(x)$ is the field basis.

\item Step 2: Feynman gates are applied during this Step.  Thus, the input states are transformed to 

\begin{equation}
	\ket{ \sum_{i=0}^{n-1}   A_i \cdot x^{2 \cdot i} \mod f(x)}
	\label{equation:GFS_DAC_7}
\end{equation}
    
Where $f(x)$ is the field basis.

\item Step 3: Step 3 of the proposed algorithm does not perform any new 
transformations on the input states.

\end{itemize}

Thus, we can see that the proposed algorithm will produce $A^2$ on the quantum 
register where $A$ is stored originally.  This proves the correctness of the 
proposed quantum Galois field squaring algorithm.

\section{Quantum Gate and Qubit Cost \\ Analysis}
\label{squarzanalytics}

The proposed quantum Galois field squaring circuits can be designed by 
following the three steps described previously in section 
\ref{algorithmmmmm}.  The quantum gate and qubit cost analysis of the squaring circuits is 
performed by analyzing the steps involved in the proposed quantum Galois field 
squaring circuit algorithm. \textit{In our work, the 
quantum gate cost is defined as the total number of Clifford+T gates in the 
circuit.  Each Clifford+T gate's cost is considered as unity.}

\begin{itemize}

\item Step 1:
\begin{itemize}
\item This Step has a qubit cost of $n$.
\item This Step has a quantum gate cost of $0$.
\end{itemize}

% of the proposed algorithm requires $n$ qubits.  Thus, this step will have a qubit cost of $n$.  No quantum gates are applied during this step.  Thus, Step 1 has a quantum gate cost of $0$.

\item Step 2:
\begin{itemize}
\item This Step does not add to the qubit cost
\item This Step has a different quantum gate cost for each Galois field. The quantum cost will depend on the result of:

\begin{equation}
	\alpha_i \cdot x^{2 \cdot i} \mod f(x)
	\label{equation:GFS_DAC_5}
\end{equation}
 
 for each input qubit $\ket{\alpha_i}$ where $\lceil \frac{n-1}{2} \rceil \leq i \leq n-1$.   $f(x)$ is the field basis and $\alpha_i \in GF(2^n)$.
 
\end{itemize} 
%  of the proposed algorithm reuses the $n$ qubits required in Step 1.  Thus, this step will have a qubit cost of $n$.  Step 2's quantum gate cost is a function of the following:

%\begin{equation}
%	\sum_{i = \lceil \frac{n-1}{2} \rceil}^{n-1} \alpha_i \cdot x^{2 \cdot i} 
%	\mod 
%	f(x)
%	\label{equation:GFS_DAC_5}
%\end{equation}

%Where $f(x)$ is the field basis and $\alpha_i \in GF(2^n)$. Thus, Step 2 will have a different quantum gate cost for each Galois field.

\item Step 3:
\begin{itemize}
\item This Step does not add to the qubit cost
\item This Step has a quantum gate cost of $0$.
\end{itemize}

% of the proposed algorithm reuses the $n$ qubits required in Step 1.  Thus, this step will have a qubit cost of $n$.  No new quantum gates are applied during this step.  Thus, Step 1 has a quantum gate cost of $0$.

\end{itemize}

Thus from Step 1, Step 2 and Step 3, the total quantum qubit cost of the 
Galois field squaring circuits from our proposed algorithm is $n$ while the 
quantum gate cost will vary. %as a function of \ref{equation:GFS_DAC_5}. 

\section{Comparison of Quantum Galois Field Squaring Circuits from our Proposed Algorithm}
\label{squarzperformance}

% we compare this work in terms of gate costs and qubit costs
\begin{table}[tbhp]

\centering
	\caption{Comparison of the quantum  circuits generated by the proposed 
	Galois field squaring algorithm against the 
	circuits in paper \cite{S1} for several field basis. }
	\resizebox{\columnwidth}{!}{	
	%{\scriptsize
	\begin{tabular}{cccccccccccc}
		\midrule
		\multicolumn{12}{c}{\textbf{Quantum Squaring Circuit Performance 
		Comparisons}} \\ \midrule
						&		
					&		
					\multicolumn{3}{c}{	\textbf{Qubits}	}		&		
					&		
					\multicolumn{3}{c}{	\textbf{Gates}	}	&	&	
					\multicolumn{2}{c}{\textbf{Depth}}	\\ \cmidrule{3-5} 
					\cmidrule{7-9} \cmidrule{11-12}
\multicolumn{1}{c}{	\textbf{Field }	}				&		&	1	
&	This	&	\% Imp. 	&		
				&	1	&	This	&	\% Imp. &	&	
				1	&	This	\\
\multicolumn{1}{c}{	\textbf{Size }	}						&		&		
& work	& w.r.t 1		
						&		&		& work	& w.r.t 1	& 
						&		&	work	\\ \toprule
		%	$	4	$	&		&	4	&	8	&	50.00	&		
		%	&	2	&	6	&	66.67	& &	1	&	2	\\
		%	$	5	$	&		&	5	&	10	&	50.00	
		%	&		&	3	&	8	&	62.50	& &	3	&	3	\\
			$	10	$	&		&	20	&	10	&	50.00 &			&	16	&	6	&	62.50	& &	4	&	2	\\
$	15	$	&		&	30	&	15	&	50.00	&		&	22	&	7	&	68.18	& &	2	&	1	\\
$	20	$	&		&	40	&	20	&	50.00	&		&	31	&	11	&	64.52	& &	4	&	2	\\
	$50 	$	&		&	100	&	50	&	50.00	&		&	129	&	79	&	38.76	& &	NA	&	6	\\
	$64 	$	&		&	128	&	64	&	50.00	&		&	165	&	101	&	38.79	& &	NA	&	7	\\
	$100 	$	&		&	200	&	100	&	50.00	&		&	264	&	164	&	37.88	& &	NA	&	8	\\
	$127 	$	&		&	254	&	127	&	50.00	&		&	190	&	63	&	66.84	& &	2	&	1	\\
	$256 	$	&		&	512	&	256	&	50.00	&		&	652	&	396	&	39.26	& &	NA	&	6	\\
	$512 	$	&		&	1024	&	512	&	50.00	&		&	1291	&	779	&	39.66	& &	NA	&	8	\\ 
	\midrule
			 \multicolumn{12}{l}{1 is the design in \cite{S1}.  Table entries marked NA where data is unavailable for design in \cite{S1}.} \\ \bottomrule
			 %\multicolumn{12}{l}{for the design in \cite{S1}.} \\ \bottomrule
			 %\multicolumn{12}{l}{2 is the proposed design.} \\ 
				
	\end{tabular}
	}
	%}
	\label{table:GFS16}
\end{table}

There are existing designs of quantum circuits for Galois field squaring such 
as the designs in \cite{S1}.  Table \ref{table:GFS16} illustrate the 
comparison between our proposed Galois field squaring circuits with those in 
\cite{S1} for several representative Galois fields.  The quantum gate cost, 
qubit cost and depth will be used for comparison.  Table \ref{table:GFS16} shows that the quantum 
Galois field squaring circuits from our algorithm excel against the existing 
Galois field squaring designs in \cite{S1} in terms of quantum gate cost, 
qubit cost and depth.  Circuit depth was reported for Galois field squaring 
designs in \cite{S1} for only Galois fields defined by irreducible 
trinomials.  From table \ref{table:GFS16} it can be seen that the 
Galois field squaring circuits from out proposed algorithm achieve improvement 
ratio of $50.00 \%$ and improvement ratios ranging from $37.88 \%$ to $68.18 \%$ 
compared to the designs in \cite{S1} in terms of qubit cost and quantum gate 
cost.

\section{Proposed Galois Field Exponentiation Algorithm}
\label{expo}

\begin{figure}
	\small
	\flushleft
	\begin{subfigure}[thb]{1.5in}
	\flushleft
		\[
		\Qcircuit @C=1em @R=.7em {
		\lstick{A} & \gate{K} & \qw & \rstick{A^2}\\
		} 
		\]
	\caption{Proposed quantum Galois field squaring circuit}
	\end{subfigure}  \qquad \begin{subfigure}[thb]{1.5in}
	\flushleft
		\[
		\Qcircuit @C=1em @R=.7em {
		\lstick{A^2} & \gate{K^{-1}} & \qw & \rstick{A}\\
		} 
		\]	
	\caption{Logical reverse of proposed quantum Galois field squaring 
	circuit}	
	\end{subfigure} \\
	\begin{subfigure}[thb]{1.5in}
		\flushleft
			\[
			\Qcircuit @C = .7em @R = .7em {
			\lstick{A} & \ctrl{1} & \qw & \rstick{A}\\
			\lstick{B} & \ctrl{1} & \qw & \rstick{B}\\
			\lstick{0} & \gate{U} & \qw & \rstick{A \cdot B} \\
			}
			\]
		\caption{Quantum Galois field multiplication circuit}
		\end{subfigure}  \qquad \begin{subfigure}[thb]{1.5in}
		\flushleft
			\[
			\Qcircuit @C = .7em @R = .7em {
			\lstick{A} & \ctrl{1} & \qw & \rstick{A}\\
			\lstick{B} & \ctrl{1} & \qw & \rstick{B}\\
			\lstick{A \cdot B} & \gate{U^{-1}} & \qw & \rstick{0} \\
			}
			\]
		\caption{Logical reverse of quantum Galois field multiplication 
		circuit}	
		\end{subfigure}
	\caption{Graphical representation of components used in proposed quantum 
	Galois field exponentiation circuit.}
\label{figure:GFS2}
\end{figure}
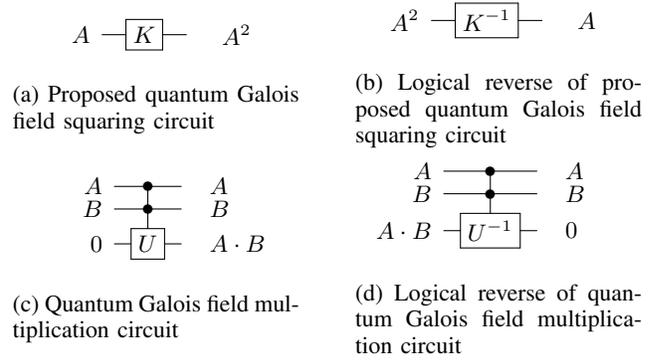

The proposed quantum Galois field exponentiation circuits is designed to minimize ancillae, is without any garbage outputs and has reduced quantum gate cost from all squaring operations.  The proposed quantum Galois field exponentiation circuit is based on the proposed quantum Galois field squaring circuit, the logical reverse of the proposed quantum Galois field squaring circuit, an existing quantum Galois field multiplication circuit and the logical reverse of the existing quantum Galois field multiplication circuit.  Figure \ref{figure:GFS2} shows the graphical representation of components used in the proposed Galois field exponentiation circuit.  

The proposed quantum Galois field squaring circuit and the logical reverse of the proposed quantum Galois field squaring circuit are 1 input and 1 output quantum circuits.  The proposed quantum Galois field squaring circuit has the mapping $A$ to $A^2$ and the logical reverse of the proposed quantum Galois field squaring circuit has the mapping $A^2$ to $A$.  Existing designs of quantum Galois field multiplication circuits such as the designs in \cite{Cheung} are 3 input and 3 output quantum circuits.  The existing designs of quantum Galois field multiplication have the mapping $A,B,C$ to $A,B, A \cdot B + C$. To implement $A \cdot B$, $C$ is replaced by ancillae.  The logical reverse of the quantum Galois field multiplication circuit is a 3 input and 3 output quantum circuit and has the mapping $A,B, A \cdot B + C$ to $A,B,C$.

Consider the exponentiation of an $n$ bit field element $a$ stored at quantum 
register $\ket{A}$.  Further consider a vector of $n-1$ quantum registers 
$\ket{B}$.  Each quantum register stores $n$ ancillae.  At the end of 
computation, quantum register $\ket{B[n-2]}$ will have $A^{2^m-2} \mod f(x)$, 
while the quantum register $\ket{A}$ keeps the value $a$.  Further, at the end 
of computation the quantum registers $\ket{B[0:n-3]}$ leave the circuit 
unchanged.  

\begin{figure*}[tbhp]
\centering
	\includegraphics[width = 6.5in]{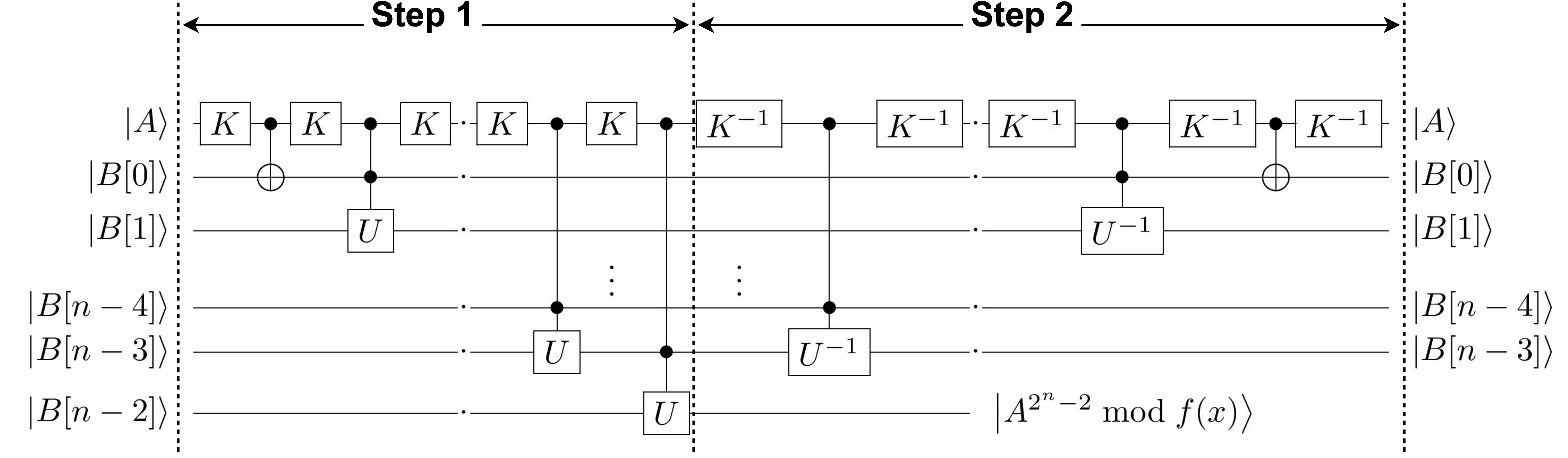}
	\caption{Circuit generation of Galois field exponentiation of an $n$ bit field value: Steps 1-2.}
\label{figure:GFS3}	
\end{figure*}

The generalized design methodology of designing the quantum Galois field exponentiation of an $n$ bit field value is shown in figure \ref{figure:GFS3}.  The generalized algorithm for designing the quantum Galois field exponentiation circuit is discussed below.  

\subsection{Steps of the Proposed Algorithm for Galois field exponentiation}

\begin{itemize}

\item  Step 1 has the following three sub-steps

\begin{itemize}

\item Step 1: At location $\ket{A}$, apply a quantum Galois fields squaring 
circuit from our proposed algorithm.

\item Step 2: For $i = 0:1:n$

Apply a Feynman gate such 
that the locations $\ket{A_i}, \ket{B[0]_i}$ are passed to the inputs $A,B$ respectively.

\item Step 3: For $i = 2:1:n-1$

At location $\ket{A}$, apply a quantum Galois fields squaring circuit from our 
proposed algorithm.

Apply a quantum Galois 
field multiplication circuit such that the locations $\ket{A},\ket{B[i-2]}$ 
and $\ket{B[i-1]}$ are passed to the inputs $A,B$ and $C$, respectively.

\end{itemize}

%Thus, after this Step, location $\ket{B[n-2]}$ is transformed to the value $A^{2^n-2} \bmod f(x)$ and the locations $\ket{B[i]}$, where $0 \leq i \leq n-3$ contain garbage outputs.  
The quantum Galois field exponentiation circuit for an $n$ bit Galois field after this Step is shown in figure \ref{figure:GFS3}.

\item Step 2 has the following four sub-steps

\begin{itemize}

\item Step 1: For $i = n-2:-1:2$

At location $\ket{A}$ apply the logical reverse of the quantum Galois field 
squaring circuit from our proposed algorithm.

Apply the logical 
reverse 
of the quantum Galois field multiplication circuit such that the locations 
$\ket{A}$, $\ket{B[i-2]}$ 
and $\ket{B[i-1]}$ are passed to the inputs $A,B$ and $C$, respectively.

\item Step 2: At location $\ket{A}$ apply the logical reverse of the quantum 
Galois field squaring circuit from our proposed algorithm.

\item Step 3: For $i = 0:1:n$

Apply a Feynman gate such 
that the locations $\ket{A_i}, \ket{B[0]_i}$ are passed to the inputs $A,B$ respectively.

\item Step 4: At location $\ket{A}$ apply the logical reverse of the quantum 
Galois field squaring circuit from our proposed algorithm.

\end{itemize} 

%Thus, after this Step, the locations $\ket{B[i]}$, where $0 \leq i \leq n-3$ are restored to their initial values. 
 The quantum Galois field exponentiation circuit for an $n$ bit Galois field after this Step is shown in figure \ref{figure:GFS3}.

\end{itemize}

\textbf{Theorem:} \textit{Let $a$ be a $n$ bit binary number represented as 
$\ket{A}$ and let $\ket{B}$ be a $n-1$ element quantum register vector of $n$ 
qubit ancillae, then the steps of the proposed Galois field exponentiation 
circuit algorithm results in a quantum Galois field exponentiation circuit 
that works correctly.  The proposed algorithm designs a quantum Galois field 
exponentiation circuits that produces the result $a^{2^m-2} \mod f(x)$ at the 
quantum register $\ket{B[n-2]}$, while the quantum register where $a$ is 
stored 
is restored to the value $a$.  Further, the quantum registers $\ket{B[0:n-3]}$ 
are restored to their original values.}

\textbf{Proof:} The proposed algorithm will make the following changes on the 
inputs: 

\begin{itemize}

\item Step 1: Step 1 of the proposed algorithm transforms the input states 
to

\begin{equation}
 \ket{A^{2^{n-1}}} \ket{A^2} \left( \bigotimes_{i = 
 2}^{n-1} \ket{\prod_{j = 1}^{i} A^{2^j}} \right) (\bmod f(x))    
	\label{equation:GFS_DAC_10}
\end{equation}

Where $f(x)$ is the field basis.

\item Step 2: Step 2 of the proposed algorithm transforms the input states 
to 

\begin{equation}
	\ket{A} \left( \bigotimes_{i=0}^{n-3} \ket{B[i]} \right) \ket{A^{2^n-2} 
	\mod f(x)}
	\label{equation:GFS_DAC_11}
\end{equation}

Where $f(x)$ is the field basis.

\end{itemize}

Thus, we can see that the proposed algorithm will produce the result at the 
quantum register where $\ket{B[n-2]}$ is stored initially, while the quantum 
register $\ket{A}$ initially holding $a$ is restored to the value $a$.  The 
quantum registers $\ket{B[0:n-3]}$ will be restored to their initial values after Step 2.  
This proves the correctness of the proposed quantum Galois field 
exponentiation algorithm.

\section{Quantum Gate and Qubit Cost \\ Analysis}
\label{expoanalytics}

\begin{table}[bthp]

\centering
	\caption{Comparison of the quantum  exponentiation circuits with the proposed Galois field squaring circuits 
	against identical Galois field exponentiation circuits using the Galois field squaring circuits in \cite{S1}.  }
	%{ \small
	\resizebox{\columnwidth}{!}{
	\begin{tabular}{ccccccccc}
		\midrule
		\multicolumn{9}{c}{\textbf{Quantum Exponentiation Circuit Performance 
		Comparisons}} \\ \midrule
					&		
					&		
					\multicolumn{3}{c}{	\textbf{Gate cost from squarings}	
					}		&		
					&		
					\multicolumn{3}{c}{	\textbf{Qubits}	}	\\ \cmidrule{3-5} 
					\cmidrule{7-9}
	\multicolumn{1}{c}{	\textbf{Field }	}				&		&	w.1 
		
	&	w. This &	\% Imp.	&		
				&	w. 1	&	w. This &	\% Imp. \\
\multicolumn{1}{c}{	\textbf{Size }	}						&		& 	& work	
&	
						w.r.t 1	
						&		&	& work	&	
						w.r.t 1	\\ 	\toprule
%$	4 $	&	&	6	&	18	&	66.67	&	&	16	&	24	&	33.33	\\
%$	5	$	&	&	12	&	32	&	62.50	&	&	25	&	40	&	37.50	\\
$	10	$	&	&	144	&	54	&	62.50	&	&	180	&	100	&	44.44	\\
$	15	$	&	&	308	&	98	&	68.18	&	&	420	&	225	&	46.43	\\
$	20	$	&	&	589	&	209	&	64.52	&	&	760	&	400	&	47.37	\\
$	50	$	&	&	6321	&	3871	&	38.76	&	&	4900	&	2500	&	48.98	\\
$	64	$	&	&	10395	&	6363	&	38.79	&	&	8064	&	4096	&	49.21	\\
$	100	$	&	&	26136	&	16236	&	37.88	&	&	19800	&	10000	&	49.49	\\
$	127	$	&	&	23940	&	7938	&	66.84	&	&	32004	&	16129	&	49.60	\\
$	256	$	&	&	166260	&	100980	&	39.26	&	&	130560	&	65536	&	49.80	\\
$	512	$	&	&	659701	&	398069	&	39.66	&	&	523264	&	262144	&	49.90	\\

 \midrule
			 \multicolumn{9}{l}{* 1 is the design in \cite{S1}} \\ \bottomrule

	\end{tabular}
	}
	\label{table:GFS17}
\end{table}

The proposed quantum Galois field exponentiation circuit can be designed with 
the two steps described in section \ref{expo}.  The quantum gate and qubit 
cost analysis of the proposed quantum Galois field exponentiation circuit are 
performed by analyzing the steps involved in the design of the proposed 
circuits.  In this analysis, $G_K$ represents the quantum gate cost of the 
proposed Galois field squaring circuits, $G_U$ represents the quantum gate 
cost of the quantum Galois field multiplier and $G_{K^{-1}}$ and $G_{U^{-1}}$ 
are the quantum gate costs for the logical reverses of the Galois field 
squaring circuits and Galois field multiplication circuits respectively.

\begin{itemize}

\item Step 1:

\begin{itemize}

\item This Step has a qubit cost of $n^2$.
\item This Step has a quantum gate cost of $(n-1) \cdot G_K + (n-2) \cdot G_U$.

\end{itemize}
% of the proposed algorithm requires $n^2$ qubits.  Thus, this step will have a qubit cost of $n^2$.  Step 1 requires $n-1$ squaring circuits and $n-2$ multiplication circuits.  Thus, this step will have a quantum gate cost of $(n-1) \cdot G_K + (n-2) \cdot G_U$.

\item Step 2:

\begin{itemize}
\item This Step does not add to the qubit cost.
\item This Step has a quantum gate cost of $(n-1) \cdot G_{K^{-1}} + (n-2) \cdot G_{U^{-1}}$.
\end{itemize}
% of the proposed algorithm reuses the $n^2$ qubits required in step 1.  Thus, this step will have a qubit cost of $n^2$.  Step 2 required $n-1$ circuits to logically reverse the squaring operation and $m-3$ circuits to logically reverse the multiplication operation.  Thus, this step will have a quantum gate cost of $(n-1) \cdot G_{K^{-1}} + (n-2) \cdot G_{U^{-1}}$.

\end{itemize}

Thus from Step 1 and Step 2, the total quantum gate cost of the quantum 
exponentiation circuit can be summed up as $(n-1) \cdot G_K + (n-2) \cdot G_U 
+ (n-1) \cdot G_{K^{-1}} + (n-2) \cdot G_{U^{-1}}$, while the qubit cost is 
$n^2$.

\section{Comparison of Quantum Galois Field Exponentiation Circuits}
\label{expoperformance}

We present the quantum circuit implementation of Galois field exponentiation 
for the first time in the literature.  Thus, we are comparing the proposed 
quantum Galois field exponentiation circuits made using our proposed Galois 
field squaring circuits with the proposed quantum exponentiation circuits made 
using the Galois field squaring circuit designs in \cite{S1} in terms of 
quantum gate costs from all squaring operations and qubit cost.  Table 
\ref{table:GFS17} illustrates the comparison of the proposed quantum Galois 
field 
exponentiation circuit with squaring circuits from our proposed algorithm with 
proposed Galois field exponentiation circuit with the existing squaring 
circuit designs 
in \cite{S1} for several representative Galois fields.  Table 
\ref{table:GFS17} shows that quantum Galois field 
exponentiation circuit with squaring circuits from our proposed algorithm 
excel against quantum exponentiation circuits made with existing squaring 
circuit designs in terms of qubit cost and the quantum gate cost  from all 
squaring operations.  From table 
\ref{table:GFS17} it can be seen that quantum Galois field exponentiation 
circuits made with squaring circuits from our proposed algorithm see 
improvement ratios ranging from $44.44 \%$ to $50.00 \%$ and $37.88 \%$ to 
$68.18 \%$ compared to quantum Galois field exponentiation circuits made with 
the squaring circuit designs in \cite{S1} in terms of qubit cost and the total 
quantum gate cost from all squaring operations.

%===========================================================

\section{Conclusion}
\label{finis}

 In this work, we have presented an algorithm to generate quantum circuits for 
 Galois field squaring using Feynman gates.  The generated quantum Galois 
 field squaring circuits are optimized for gate cost, depth and have zero 
 overhead in terms of number of ancillae compared to existing designs.  
 Further, we present an algorithm for the quantum circuit implementation of 
 Galois field exponentiation that uses proposed quantum squaring circuit.  
 The proposed algorithm to generate quantum circuits for Galois 
 field squaring is verified by formal proof and with functional verification 
 of generated Galois field squaring circuits in MATLAB.  The proposed algorithm to generate quantum circuits for Galois 
 field exponentiation is verified by formal proof.  The Galois field 
 squaring circuits and Galois field exponentiation circuits generated from our 
 proposed algorithms will find promising 
 applications in quantum computing and could form components of quantum Galois 
 field processing architectures.

\bibliographystyle{IEEEtran}
\bibliography{IEEEabrv,GSbiblio.bib}
%\begin{thebibliography}{1}
%\end{thebibliography}

\end{document}